\documentclass[sigconf]{acmart}
\AtBeginDocument{%
  \providecommand\BibTeX{{%
    \normalfont B\kern-0.5em{\scshape i\kern-0.25em b}\kern-0.8em\TeX}}}

\usepackage{amsmath,amsfonts,bm}

\def\eqref#1{equation~\ref{#1}}

\def\1{\bm{1}}

\DeclareMathAlphabet{\mathsfit}{\encodingdefault}{\sfdefault}{m}{sl}
\SetMathAlphabet{\mathsfit}{bold}{\encodingdefault}{\sfdefault}{bx}{n}

\DeclareMathOperator*{\argmax}{arg\,max}

\usepackage{hyperref}
\usepackage{url}
\usepackage{soul}
\usepackage[normalem]{ulem}
\usepackage{wrapfig,lipsum,booktabs}
\usepackage{hyperref}
\usepackage{url}
\usepackage{dsfont}
\usepackage{booktabs} 
\usepackage{caption}
\usepackage{float}
\usepackage{amsmath}
\usepackage{mathtools}
\usepackage{sidecap}
\usepackage{multirow}
\usepackage{epsfig,graphics}

\usepackage{todonotes}
\usepackage[ruled,vlined]{algorithm2e}
\usepackage{subfig,balance,lineno}
\usepackage{verbatim}

\usepackage{wrapfig}

\copyrightyear{2021}
\acmYear{2021}
\setcopyright{rightsretained}
\acmConference[KDD '21]{Proceedings of the 27th ACM SIGKDD
Conference on Knowledge Discovery and Data Mining}{August 14--18,
2021}{Virtual Event, Singapore}
\acmBooktitle{Proceedings of the 27th ACM SIGKDD Conference on
Knowledge Discovery and Data Mining (KDD '21), August 14--18, 2021,
Virtual Event, Singapore}\acmDOI{10.1145/3447548.3467153}
\acmISBN{978-1-4503-8332-5/21/08}

\begin{document}


\title{Multimodal Emergent Fake News Detection via \\Meta Neural Process Networks}


\author{Yaqing Wang$^{\S}$, Fenglong Ma$^\diamond$, Haoyu Wang$^{\S}$, Kishlay Jha$^\dagger$ and Jing Gao$^{\S}$}
\affiliation{\institution{{$^{\S}$}Purdue University, West Lafayette, Indiana, USA\\
$^\diamond$Pennsylvania State University, Pennsylvania, USA\\
$^\dagger$University of Virginia, Charlottesville, Virginia, USA\\
$^{\S}$\{wang5075,  jinggao, wang5346\}@purdue.edu, 
$^\diamond$fenglong@psu.edu,
$^\dagger$kishlay@email.virginia.edu
}
\country{}
}








\begin{abstract}
  
Fake news travels at unprecedented speeds, reaches global audiences and puts users and communities at great risk via social media platforms. Deep learning based models show good performance when trained on large amounts of labeled data on events of interest, whereas the performance of models tends to degrade on other events due to domain shift. Therefore, significant challenges are posed for existing detection approaches to detect fake news on emergent events, where large-scale labeled datasets are difficult to obtain. Moreover, adding the knowledge from newly emergent events requires to build a new model from scratch or continue to fine-tune the model, which can be challenging, expensive, and unrealistic for real-world settings.  In order to address those challenges, we propose an end-to-end fake news detection framework named MetaFEND, which is able to learn quickly to detect fake news on emergent events with a few verified posts. Specifically, the proposed model integrates meta-learning and neural process methods together to enjoy the benefits of these approaches.  In particular, a label embedding module and a hard attention mechanism are proposed to enhance the effectiveness by handling categorical information and trimming irrelevant posts. Extensive experiments are conducted on multimedia datasets collected from Twitter and Weibo. The experimental results show our proposed MetaFEND model can detect fake news on never-seen events effectively and outperform the state-of-the-art methods.

\end{abstract}



\fancyhead{}
%

\maketitle

{\fontsize{8pt}{8pt} \selectfont
\textbf{ACM Reference Format:}\\
Yaqing Wang, Fenglong Ma, Haoyu Wang, Kishlay Jha, Jing Gao. 2021. Multimodal Emergent Fake News Detection via Meta Neural Process Networks. In {\it Proceedings of the 27th ACM SIGKDD Conference on Knowledge Discovery and Data Mining (KDD'21), August 14--18, 2021, Virtual Event, Singapore.} ACM, New York, NY, USA, 9 pages. https://doi.org/10.1145/3447548.3467153 }

\section{Introduction}
The recent proliferation of social media has significantly changed the way in which people acquire information.
According to the 2018 Pew Research Center survey, about two-thirds of American adults (68\%) get news on social media at least occasionally.  The fake news on social media usually take advantage of multimedia content which contain misrepresented or even forged images, to mislead the readers and get rapid dissemination. The dissemination of fake news may cause large-scale negative effects, and sometimes can affect or even manipulate important public events. Recent years have witnessed a number of high-impact fake news spread regarding terrorist plots and attacks, presidential election and various natural disasters. 
Therefore, there is an urgent need for the development of automatic detection algorithms, which can detect fake news as early as possible to stop the spread of fake news and mitigate its serious negative effects.

\begin{figure}[hbt]
\centering
\vspace{-0.15in}
\includegraphics[width=3.2in]{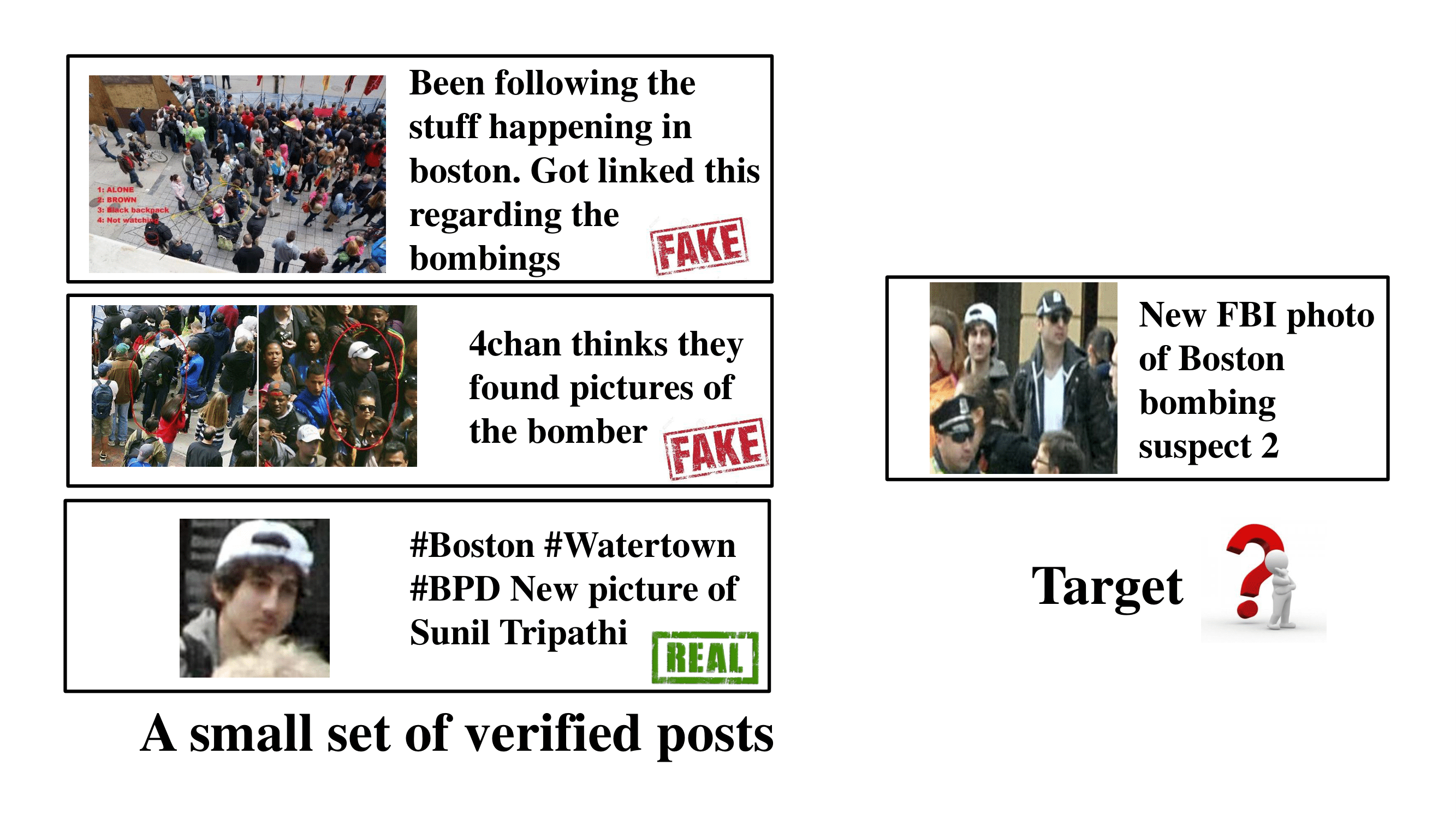}
\vspace{-0.25in}
 \caption{Fake news examples on an emergent event Boston Bombing  from Twitter.}\label{Fig:example}
\vspace{-0.1in}
\end{figure}
\noindent\textbf{Task Challenges.} Thus far, various fake news detection methods, including both traditional learning~\citep{conroy2015automatic, tacchini2017some} and deep learning based models~\citep{ruchansky2017csi,ma2016detecting,ma2018detect, ma2019detect, EANN,popat2018declare} have been exploited to identify fake news. Despite the success of  deep learning models with large amounts of labeled datasets,  the algorithms 
still suffer in the cases where  fake news detection is needed on emergent events. Due to the domain shift in the news events~\cite{WeFEND}, the model trained on past events may not achieve satisfactory performance and thus the new knowledge from emergent events are needed to add into fake news detection models. However, adding the knowledge from newly emergent events requires to build a new model from scratch or continue to fine-tune the model on newly collected labeled data, which can be challenging, expensive, and unrealistic for real-world settings. Moreover, fake news usually emerged on newly arrived events where we hardly obtain sufficient posts in a timely manner. In the early stage of emergent events,  we usually only have a handful of related verified posts (An example is shown in the Fig.~\ref{Fig:example}). How to leverage \emph{a small set of verified posts} to make the model learn quickly to detect fake news on the newly-arrived events is a crucial challenge.

\noindent\textbf{Limitations of Current Techniques.}  To overcome the challenge above, the few-shot learning, which aims to leverage a small set of data instances for quick learning, is a possible solution. One promising research line of few-shot learning is \textbf{meta-learning}~\citep{MAML, MetaSGD}, whose basic idea is to leverage the global knowledge from previous tasks to facilitate the learning on new task. However,  the success of existing meta-learning methods is highly associated with an important assumption: the tasks are from a similar distribution and the shared global knowledge applies to different tasks. This assumption usually does not hold in the fake news detection problem as the writing style, content, vocabularies and even class distributions of news on different events usually tends to differ. 
\begin{figure}[hbt]
\vspace{-0.2in}
\subfloat[Twitter]{
\begin{minipage}{.23\textwidth}
\includegraphics[height=1.2in]{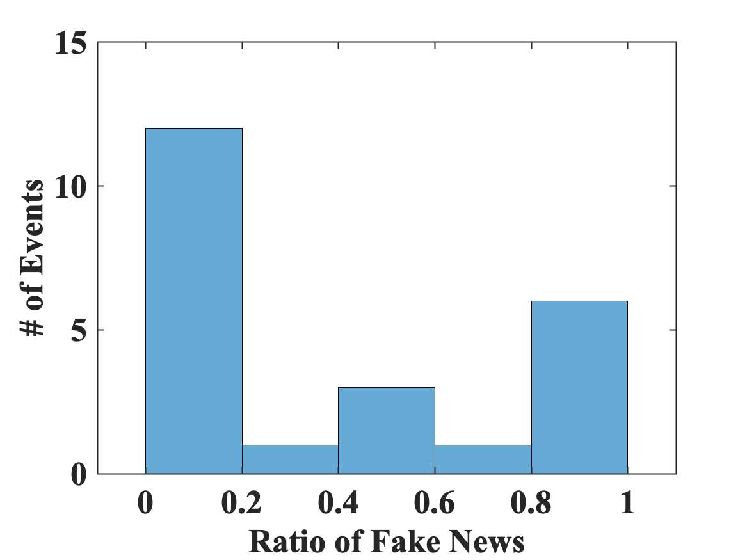}
\label{subfig:twitter}
\end{minipage}}
\subfloat[Weibo]{
\begin{minipage}{.23\textwidth}
\includegraphics[height=1.2in]{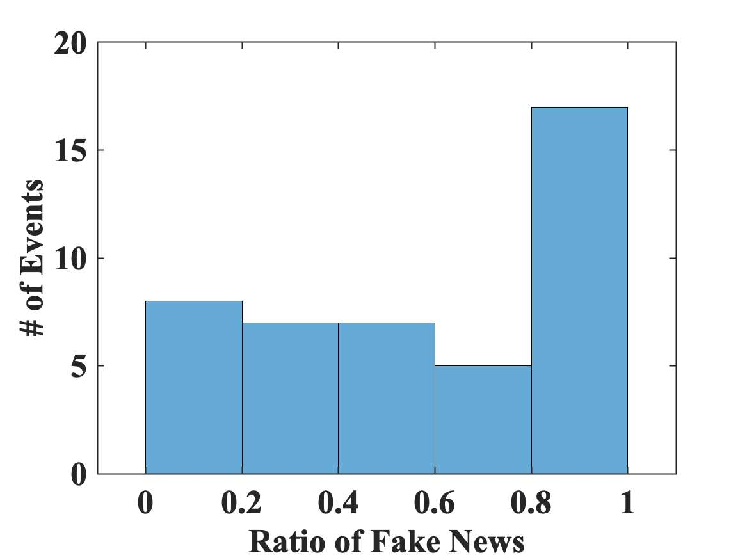}
\label{subfig:weibo}
\end{minipage}}
 \caption{The number of events with respect to different percentages of fake news.}\label{fig:histogram}
\vspace{-0.1in}
\end{figure}
As it can be observed from Figure~\ref{fig:histogram}, the ratios of fake news on events are significantly different. The significant difference across events posts serious challenges on \textbf{event heterogeneity}, which cannot be simply handled by globally sharing knowledge~\citep{hsml}. 
Another research line of few-shot learning is \textbf{neural processes}~\citep{cnp, np, anp}, which conduct inference using a small set of data instances as conditioning. Even though neural processes show better generalizablity , they are based on a fixed set of parameters and usually suffer from the limitations like \textbf{underfitting}~\citep{anp}, thereby leading to unsatisfactory performance.  These two research lines of models are complementary to each other: the parameter adaptation mechanism in meta-learning can provide more parameter flexibility to \emph{alleviate unfitting issues} of the neural process. Correspondingly, the neural processes can help handle \emph{the heterogeneity challenge} for MAML by using a small set of data instances as conditioning instead of encoding all the information into parameter set. Although it is promising to integrate two popular few-shot approaches together, the incompatible operations on the given small set of data instances is the main obstacle for developing the model based on these two.

\noindent\textbf{Our Approach.} To address the aforementioned challenges, in this paper, we propose a novel meta neural process network (namely MetaFEND) for emergent fake news detection. MetaFEND unifies the incompatible operations from meta-learning and neural process via a simple yet novel simulated learning task, whose goal  is to  adapt the parameters to better take advantage of given support data points as conditioning. Toward this end, we propose to conduct leave-one-out prediction as shown in the Fig.~\ref{Fig:framework}, i.e., we repeatedly use one of given data as target data and the rest are used as context set for conditioning on all the data in support set. Therefore, the proposed model can handle heterogeneous events via event adaption parameters and conditioning on event-specific data instances simultaneously. Furthermore, we incorporate two novel components - \emph{label embedding} and \emph{hard attention} - to handle categorical characteristics of label information and extract the most informative instance as conditioning despite  imbalanced class distributions of news events.  Experimental results on two large real-world datasets show that the proposed model effectively detect fake news on new events with a handful of posts  and outperforms the state-of-the-art approaches.

\noindent\textbf{Our Contributions.} The main contributions of this paper can be summarized as follows:
\begin{itemize}
\item We recognize the challenges of fake news detection on emergent events and formulate the problem into a few-shot learning setting. Towards this end, we propose an effective meta neural process framework to detect fake news on emergent events with a handful of data instances.

\item The proposed MetaFEND method fuses the  meta-learning method and neural process models together via a simulated learning task design. We also propose two components  \emph{label embedding} and  \emph{hard attention} to handle categorical information and select the formative instance respectively. The effects of two components are investigated in the experiments.

\item We empirically show that the proposed method MetaFEND can effectively identify fake news on various events and largely outperform the state-of-the-art models on two real-world datasets.
\end{itemize} 
\section{Background}
We define our problem and introduce preliminary works in this section.
\vspace{-0.05in}
\subsection{Problem Formulation}
\label{section:backgrond}
There are many tasks related to fake news detection, such as rumor detection~\cite{jin2014news} and spam detection~\cite{shen2017discovering}.
Following the previous work~\cite{ruchansky2017csi,shu2017fake}, we specify the definition of fake news as news which is intentionally fabricated and can be verified as false. In this paper, we tackle fake news detection on emergent events and make a practical assumption that a few labeled examples are available per event. Our goal is to leverage the knowledge learned from past events to conduct effective fake news detection on newly arrived events with a few examples. More formally, we define the fake news detection following the few-shot problem.

\noindent \textbf{Few-shot Fake News Detection}
Let $\mathcal{E}$ denote a set of news events. In each news event $e \sim \mathcal{E}$, we have {a few labeled posts} on the event $e$. The core idea of few-shot learning is to use episodic classification paradigm to simulate few-shot settings during model training. In each episode during the training stage, the labeled posts are partitioned into two independent sets, support set and query set. Let $\{\mathbf{X}_{e}^s, \mathbf{Y}_{e}^s\} = \{x^s_{e,i}, y^s_{e,i}\}^{K}_{i=1}$ represent the support set, and $\{\mathbf{X}_{e}^q, \mathbf{Y}_{e}^q\} = \{x^q_{e,i}, y^q_{e,i}\}_{i=K+1}^{N}$ be the query set. The model is trained to learn to conduct fake news detection on the query set $\{\mathbf{X}_{e}^q, \mathbf{Y}_{e}^q\}$ given the support set $\{\mathbf{X}_{e}^s, \mathbf{Y}_{e}^s\}$.  During the inference stage,  $K$ labeled posts are provided per event. For each event $e$,  the model leverages its corresponding $K$ labeled posts as support set $\{\mathbf{X}_{e}^s, \mathbf{Y}_{e}^s\} = \{x^s_{e,i}, y^s_{e,i}\}^{K}_{i=1}$ to conduct fake news detection on given event $e$. 
\subsection{Preliminary Work}

\begin{figure*}[hbt]

\centering
\includegraphics[width=5.in]{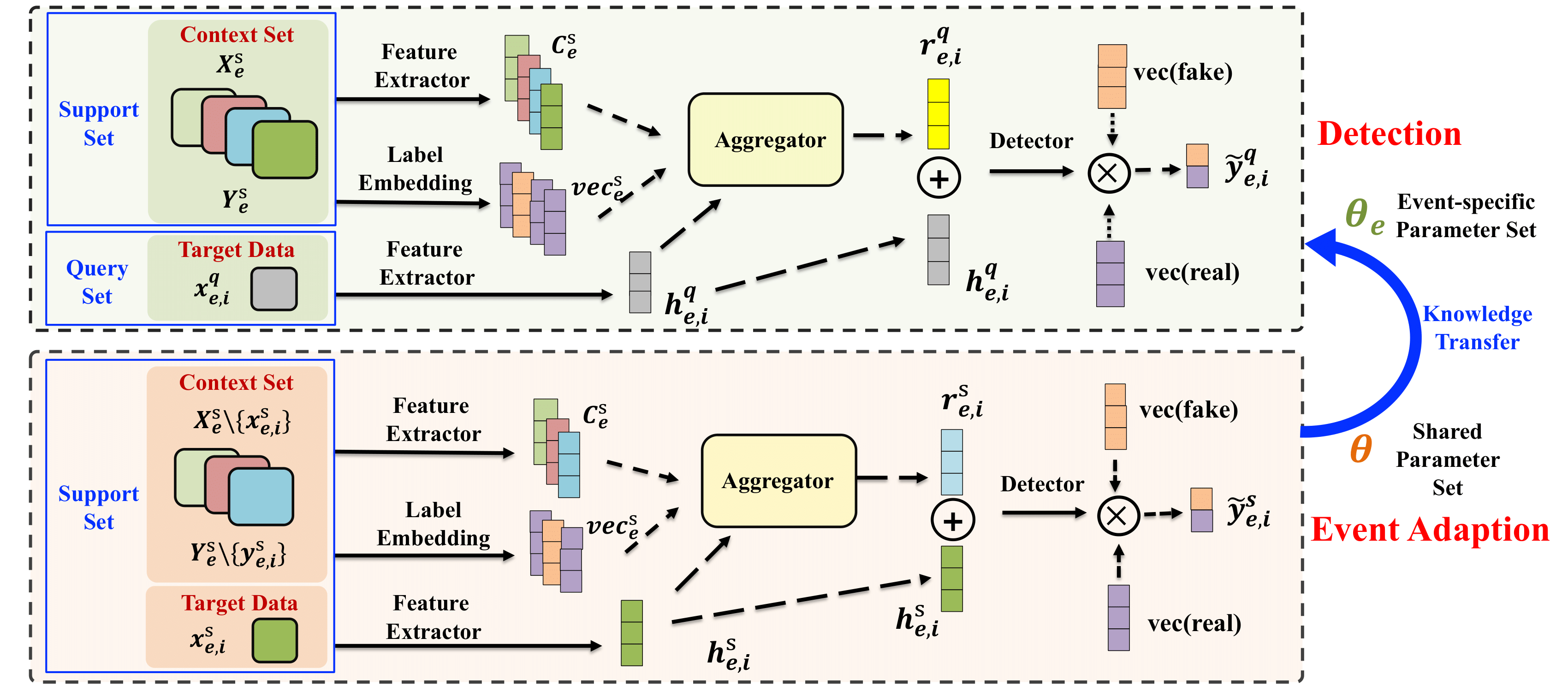}
 \caption{The proposed framework MetaFEND. The proposed framework has two stages: event adaption and detection. During the event adaption stage, the model parameter set $\theta$ is updated to event-specific parameter set  $\theta_e$. During the detection stage, the event-specific parameter set $\theta_e$ is used to detect fake news on event $e$. $\oplus$  denotes concatenation operation and $\otimes$ means element-wise product. }\label{Fig:framework}
\vspace{-0.1in}
\end{figure*}

\textbf{MAML.} We first give an overview of MAML method~\cite{MAML}, a representative algorithm of gradient-based meta-learning approaches, and take few-shot fake news detection as an example. The meta-learning procedure is split into two stages: meta-training and meta-testing.

During the {\em meta-training} stage, the baseline learner $f_{\theta}$ is adapted to specific event $e$ as $f_{\theta_e}$ with the help of the support set $\{\mathbf{X}_{e}^s, \mathbf{Y}_{e}^s\}$. Such an event specific learner $f_{\theta_e}$ is evaluated on the corresponding query set $\{\mathbf{X}_{e}^q, \mathbf{Y}_{e}^q\}$. The loss $\mathcal{L}(f_{\theta_e}, \{\mathbf{X}_{e}^q, \mathbf{Y}_{e}^q\})$ on $\{\mathbf{X}_{e}^q, \mathbf{Y}_{e}^q\}$ is used to update the parameters of baseline learner $\theta$. During the meta-testing stage, the baseline learner $f_{\theta}$ is adapted to the testing event $e'$ using the procedure in meta-training stage to obtain event specific parameters $\theta_{e'}$, which is employed to make predictions on the query set  $\{\mathbf{X}_{e'}^q, \mathbf{Y}_{e'}^q\}$ of event $e'$.  

MAML update parameter vector $\theta$  using one or more gradient descent updates on event $e$. For example, when using one gradient update:
$$
\theta_e = M(f_{\theta}, \{\mathbf{X}_{e}^s, \mathbf{Y}_{e}^s\}) =  \theta - \alpha \bigtriangledown_{\theta} \mathcal{L}(f_{\theta}, \{\mathbf{X}_{e}^s, \mathbf{Y}_{e}^s\}).
$$
The model parameters are trained by optimizing for the performance of $f_{\theta_e}$ with respect to $\theta$ across events sampled from $p(\mathcal{E})$.  More concretely, the meta-objective is as follows:
$$
\min_{\theta} \sum_{e \sim \mathcal{E}} \mathcal{L}(f_{\theta_i}) =  \sum_{e \sim \mathcal{E}} \mathcal{L}(f_{\theta - \alpha \bigtriangledown_{\theta} \mathcal{L}(f_{\theta}, \{\mathbf{X}_{e}^s, \mathbf{Y}_{e}^s\})}, \{\mathbf{X}_{e}^q, \mathbf{Y}_{e}^q\}) .
$$
\textbf{Limitations of MAML.} The MAML can capture task uncertainty via one or several gradient updates. However, in fake news detection problem, when events are heterogeneous, the event uncertainty is difficult to encode into parameters via one or several gradient steps.  Moreover, even if given support data and query data of interest are from the same event, there is no guarantee that they are all highly related to each other. In such a case, the parameter adaption on fake news detection loss on support set may be misleading for some posts.

\noindent\textbf{Conditional Neural Process (CNP).}
 The CNP includes four components: encoder, feature extractor, aggregator and decoder. The basic idea of conditional neural process is to make predictions with the help of support set $\{\mathbf{X}_{e}^s, \mathbf{Y}_{e}^s\} = \{x^s_{e,i}, y^s_{e,i}\}^{K}_{i=1}$ as context.
 The dependence of a CNP on the support set is parametrized by a neural network encoder, denoted as $g(\cdot)$.  The encoder $g(\cdot)$ embeds each observation in the support set into feature vector, and the aggregator $\mathrm{agg}(\cdot)$ maps these feature vectors into an embedding of fixed dimension.  In CNP, the aggregation procedure is a permutation-invariant operator like averaging or summation. The query data of interest $x^q_{e,i}$ is fed into feature extractor $h(\cdot)$ to get the feature vector. Then the decoder $f(\cdot)$ takes the concatenation of aggregated embedding and given target data $x_{e,i}^{q}$ as input and output the corresponding prediction as follows:
$$
p(y^q_{e,i}| \{\mathbf{X}_{e}^s, \mathbf{Y}_{e}^s\}, x_{e,i}^q) = f\big(\mathrm{agg}(g(\{\mathbf{X}_{e}^s, \mathbf{Y}_{e}^s\}))\oplus h(x_{e, i}^{q})\big).
$$
where $\oplus$ is concatenation operator.\\
\textbf{Limitations of CNP.} One widely recognized limitation of CNP is underfitting~\cite{anp}. For different context data points, their importance is usually different in the prediction. However, the aggregator of CNP treats all the support data equally and cannot achieve query-dependent context information. Moreover, the CNP simply concatenates the input features and numerical label values of posts together as input, ignoring the categorical characteristics of labels.

\section{Methodology} 

In this paper, we study how to develop an effective model which can identify fake news on emergent events with a small set of labeled data. To this end, we propose a meta neural process framework which can fuse meta-learning and neural process methods together via a simulated task. To tackle the challenges brought by heterogeneous news events,   we further propose a label embedding component to handle categorical labels and a hard attention component, which can select the most informative information from the support set with imbalanced class distributions.
In the next subsection, we introduce our overall design and architecture.

\subsection{Meta-learning Neural Process Design}

As shown in Figure~\ref{Fig:framework}, our proposed framework includes two stages: event adaptation and detection.  The event adaptation stage is to adapt the model parameters to specific event with the help of the support set. The detection stage is to detect fake news on the given event with the help of the support and the adapted parameter set. 

\noindent\textbf{Event adaption.} We take the $i$-th support data $\{x^s_{e, i}, y^s_{e,i}\}$ as an example, in the {event adaption stage}, the $\{x^s_{e, i}, y^s_{e,i}\}$ is used as target data and the rest of support set $\{\mathbf{X}^s_{e}, \mathbf{Y}^s_{e}\} \setminus \{x^s_{e, i}, y^s_{e,i}\}$ are used as context set accordingly. The context set $\{\mathbf{X}^s_{e}, \mathbf{Y}^s_{e}\} \setminus \{x^s_{e, i}, y^s_{e,i}\}$ and target data $x^s_{e, i}$ are fed into the proposed model to output the prediction.
The loss can be calculated between the prediction $\hat{y}^s_{e,i}$ and the corresponding label $y^s_{e,i}$.
For simplicity, we use $\theta$ to represent all the parameters included in the proposed model. Then, our event adaption objective function on the support set can be represented as follows:

\begin{equation} 
\label{eq:event_adaption}
 \mathcal{L}_e^s = \sum_i \log p_{\theta}(y^s_{e,i}| \{\mathbf{X}_{e}^s, \mathbf{Y}_{e}^s\} \setminus \{x^s_{e, i}, y^s_{e,i}\}, x_{e,i}^s).
\end{equation}

\noindent We then update parameters $\theta$ one or more gradient descent updates on $\mathcal{L}_e^s$ for event $e$. For example, when using one gradient update:
\begin{equation}
\label{eq:adaptation}
\theta_e = \theta - \alpha \bigtriangledown_{\theta} \mathcal{L}_e^s.
\end{equation}

\noindent\textbf{Detection stage.} The proposed model with event-specific parameter set $\theta_e$ takes query set $\mathbf{X}_e^q$ and entire support set $\{\mathbf{X}_e^s,\mathbf{Y}_e^s\}$ as input and outputs predictions $\hat{\mathbf{Y}}^q_{e}$ for  query set $\mathbf{X}_e^q$. The corresponding loss function in the detection stage can be represented as follows: 
\begin{equation}
\mathcal{L}_e^q = \log p_{\theta_e}(Y^q_e| X_e^s, Y_e^s, X_e^q). 
\end{equation}

Through this meta neural process, we can learn an initialization parameter set which can rapidly learn to use given context input-outputs as conditioning to detect fake news on newly arrived events. 

\noindent\textbf{Neural Network Architecture}. From Figure~\ref{Fig:framework}, we can observe that the network structures used in these two stages are the same, including feature extractor, label embedding, aggregator and detector. The feature extractor is a basic module which can take posts as input and output corresponding feature vectors.  Label embedding component is to capture semantic meanings of labels. Then we use an aggregator to  aggregate these information into a fixed dimensional vector, namely context embedding, which is used as reference for fake news detection. Thereafter both the context embedding and target feature vector are fed into detector to output a vector. The final prediction is based on the similarities between this output vector and label embeddings. In the following subsections, we use event adaption to introduce the details of each component in our proposed model. For simplicity, we omitted superscript $s$ and $q$ in the illustrations about components. 

\vspace{-0.1in}
\subsection{Feature Extractor}

From Figure~\ref{Fig:framework}, we can observe that feature extractor is a basic module to process raw input. Following the prior works~\cite{EANN,WeFEND},  our feature extractor consists of two parts: {textual feature extractor} and {visual feature extractor}. For a minor note, the feature extractor is a plug-in component which can be easily replaced by other state-of-the-art models.

\noindent\textbf{Textual feature extractor.} We adopt convolutional neural network~\citep{textcnn}, which is proven effective in the fake news detection~\citep{EANN,WeFEND}, as textual feature extractor. The input of the textual feature extractor is unstructured news content, which can be represented as a sequential list of words. For the $t$-th word in the sentence, we represent it by the word embedding vector which is the input to the convolutional neural network.  After the convolutions neural network, we feed the output into a fully connected layer to adjust the dimension to $d_{f}$ dimensional textual feature vector. 

\noindent\textbf{Visual feature extractor.} The attached images of the posts are inputs to the visual feature extractor. In order to efficiently extract visual features, we employ the pretrained VGG19~\citep{vgg19} which is used in the multi-modal fake news works~\cite{EANN, jin2017multimodal}. On top of the last layer of VGG19 network, we add a fully connected layer to adjust the dimension of final visual feature representation to the same dimension of textual feature vector $d_{f}$. During the joint training process with the textual feature extractor, we freeze the parameters of pre-trained VGG19 neural network to avoid overfitting.

For a multimedia post, we feed the text and image of the example into textual and visual feature extractor respectively. The output of two feature extractors are concatenated together to form a feature vector. For the target data $x_{e,i}$, we denote its feature vector as $\mathbf{h}_{e,i}$. For the context data $x_{e,k}$ where $k \neq i$, we denote its feature vector as  $\mathbf{c}_{e,k} \in \mathbf{C}_e$.

\subsection{Aggregator}
\label{section:hard_attention}

To construct context embedding for target data, we need to design an aggregator which satisfies two properties: permutation-invariant and target-dependent. To satisfy the two properties, we choose to adopt the attention mechanism which can compute weights of each observations in context set with respect to the target and aggregates the values according to their weights to form the new value accordingly.

\noindent\textbf{Attention mechanism.} In this paper, we use scaled dot-product attention mechanism~\citep{Transformer}. This attention function can be described as mapping a query and a set of key-value pairs to an output,
where the query $\mathbf{Q}$, keys $\mathbf{K}$, values $\mathbf{V}$, and output are all vectors.  In our problem, for the target data $x_{e,i}$ and the context set $\mathbf{X}_{e} \setminus \{{x}_{e,i}\} = \{x_{e, k}\}_{k=1, k \neq i}^K$ on event $e$. We use the the target feature vector $\mathbf{h}_{e,i} \in \mathbb{R}^{1 \times d}$ after linear transformation as query vector $\mathbf{Q}_{i}$, the context feature vector $\mathbf{C}_e =[c_{e,1},..., c_{e, K}] \in \mathbb{R}^{K \times d}$ after linear transformation as the Key vector $\mathbf{K}$. For the context set, we represent its label information $\mathbf{Y}_e \setminus \{{y}_{e,i}\}= \{y_{e,k}\}_{k=1, k \neq i}^K$ by semantic embeddings as $\mathbf{vec}_e =\{\mathbf{vec}_{e,k}\}_{k=1, k \neq i}^K $. The details of label embedding are introduced in the next subsection. Then we concatenate context feature vector and label embedding as $\mathbf{C_e}\oplus \mathbf{vec}_e = [\mathbf{c}_{e, 1}\oplus \mathbf{vec}_{e,1},..., \mathbf{c}_{e, K}\oplus \mathbf{vec}_{e, K}] \in \mathbb{R}^{(K-1) \times 2d}$. The concatenated embedding after linear transformation is used as value vector $\mathbf{V}$. We represent $\mathbf{Q}_{i}, \mathbf{V}, \mathbf{K}$ as follows:
$$
    \mathbf{Q}_{i} =\mathbf{W}_q \mathbf{h}_{e,i},
$$
$$
    \mathbf{K} =\mathbf{W}_k \mathbf{C}_e,
$$
$$
    \mathbf{V} =\mathbf{W}_v (\mathbf{C}_e\oplus \mathbf{vec}_e), 
$$
where $\mathbf{W}_q \in \mathbb{R}^{d \times d}$, $\mathbf{W}_k \in \mathbb{R}^{d \times d}$ and $\mathbf{W}_v \in \mathbb{R}^{2d \times d}$.

The output is computed as a weighted sum
of the values, where the weight assigned to each value is computed by dot-product function of the
query with the corresponding key.   More specifically, attention function can be represented as follows:

\begin{equation}
\mathbf{a}_i = \mathrm{softmax}(\frac{\mathbf{Q}_{i}\mathbf{K}^T}{\sqrt{d}})
\label{eq:attention}
\end{equation}
\begin{equation}
\mathrm{Attention}(\mathbf{Q}_{i}, \mathbf{K}, \mathbf{V}) := a_i \mathbf{V}.
\end{equation}

\noindent\textbf{Limitation of Soft-Attention.} The attention mechanism with soft weight values is categorized into soft-attention. However, soft-attention cannot effectively trim irrelevant data  especially when we have a context set with an imbalanced class distribution shown in Fig.~\ref{fig:histogram}. Moreover, we show a case study in the experimental section for a better illustration.

\noindent\textbf{Hard-Attention.} To overcome the limitation of soft-attention, we propose to select the most related context data point instead of using weighted average. To enable argmax operation to be differentiable, we use Straight-Through (ST) Gumbel SoftMax~\citep{gumbel} for discretely sampling the context information given target data.
 We introduce the sampling and $\argmax$ approximations of ST Gumbel SoftMax procedure next.

The Gumbel-Max trick~\cite{gumbel1948statistical} provides a simple and efficient way
to draw samples z from a categorical distribution with class probabilities. In our problem, for the $i$-th target data point $x_{e, i}$ with context set $\mathbf{X}_e  \setminus\{{x}_{e,i}\} = \{x_{e, k}\}_{k=1, k \neq i}^K$, the class probabilities can be obtained from the weight vector $\mathbf{a}_{i} = [a_{i,1},..., a_{i, K}]$ from dot-product attention mechanism according to Eq.~\ref{eq:attention}. Because $\argmax$ operation is not differentiable, we use the softmax function as a continuous, differentiable approximation to $\argmax$, and generate K-dimensional sample vectors $\mathbf{P}_{i} = [p_{i,1}, p_{i,2}.., p_{i,K}]$ as follows:
\begin{equation}
  p_{i,k}  =\frac{ \exp((\log(a_{i,k}) + g)/\tau)}{\sum_{k, k \neq i}^K \exp((\log(a_{i,k}) + g)/\tau)}
  \label{eq:gumbel_softamx}
\end{equation}

where  $\tau$ is a temperature parameter, $g = - \log( -\log(\mu))$ is the Gumbel noise and $\mu$ is generated by a certain noise distribution (e.g., $u \sim \mathcal{N}(0, 1)$). As the softmax temperature $\tau$ approaches 0,  the Gumbel-Softmax distribution becomes identical to the
categorical distribution.  Moreover, Straight-Through (ST) gumbel-Softmax takes different paths in the forward and backward propagation, so as to maintain sparsity yet support stochastic gradient descent. Through gumbel-softmax, the hard-attention mechanism is able to draw the most informative sample based on weight vectors from $\mathbf{P}_{i}$ for given target sample ${x}_{e,i}$.

 The hard-attention can trim the irrelevant data points and select the most related data point, denoted as  $\mathbf{c}_{e,k}\oplus\mathbf{v}_{e,k} \in \mathbb{R}^{2d}$. Besides the hard-attention mechanism, the aggregator includes an additional fully connected layer on top of hard-attention to adjust the dimension. The $\mathbf{c}_{e,k}\oplus\mathbf{v}_{e,k}$ is fed into this fully connected layer  to  output context embedding $\mathbf{r}_{e,i} \in \mathbb{R}^d$.

\subsection{Detector based on Label Embedding}

 \noindent\textbf{Categorical characteristic of label information.} The context information includes posts and their corresponding labels. The existing works like CNP~\citep{cnp} and ANP~\citep{anp} usually simply concatenate the input features and numerical label values together as input to learn a context embedding via a neural network. Such operation discards the fact that label variables are categorical. Moreover, this operation tends to underestimate the importance of labels as the dimension of input features is usually significantly larger than that of single dimensional numerical value. To handle categorical characteristic,  we propose to embed labels into fixed dimension vectors inspired by word embedding~\citep{w2v}. We define two embeddings $\mathbf{vec(fake)}$ and $\mathbf{vec(real)}$ for the labels of fake news and real news respectively.  For example, given the $k$-th post $x_{e,k}$ on event $e$, the corresponding label is fake and its label embedding vector is $\mathbf{vec(fake)}$, and we denote the label embedding of $x_{e,k}$ as $\mathbf{vec}_{e,k}$. To ensure that the label embedding can capture the semantic meanings of corresponding labels, we propose to use embeddings $\mathbf{vec(fake)}$ and $\mathbf{vec(real)}$ in the detector as metrics and output predictions are determined based on metric matching.

The detector is a fully-connected layer which takes target feature vector  and context embedding   as inputs and outputs a vector that has the same dimensionality as that of the label embedding. More specifically, for $i$-th target data, the context embedding $\mathbf{r}_{e,i} $ and target feature vector $\mathbf{h}_{e,i}$ are concatenated. Then the detector takes  $\mathbf{r}_{e,i} \oplus \mathbf{h}_{e,i} \in \mathbb{R}^{2d}$ as input and produces a output vector $\mathbf{o}_{e,i} \in  \mathbb{R}^{d}$. The similarities between output $\mathbf{o}_{e,i}$ from our model and label embeddings $\mathbf{vec(fake)}$ and $\mathbf{vec(real)}$  are calculated as follows:
\begin{equation}
\mathrm{similarity} (\mathbf{o}_{e,i}, \mathbf{vec(fake)}) =  \left\|\mathbf{o}_{e,i}\circ \mathbf{vec(fake)}  \right\|,
\label{eq:similiraty_fake}
\end{equation}
\begin{equation}
\mathrm{similarity} (\mathbf{o}_{e,i}, \mathbf{vec(real)}) =  \left\|\mathbf{o}_{e,i}\circ \mathbf{vec(real)}  \right\|.
\label{eq:similiraty_real}
\end{equation}
The two similarity scores are then mapped into $[0, 1]$ as probabilities via \textit{softmax}. The trainable label embedding capture semantic meaning of labels and can generalize easily to new events with the help of adaptation step according to Eq.~\ref{eq:adaptation}.

\subsection{Algorithm Flow}
\label{section:model_architecture}

After introducing the meta-learning neural process design, feature extractor, label embedding, aggregator and detector components, we  present our algorithm flow.

 As it  can be observed from Figure~\ref{Fig:framework}, when tackling an event $e$, our proposed framework MetaFEND has two stages: event adaption and detection. In more details, our proposed model adapts to the specific event according to Eq.~\ref{eq:adaptation} and then the event-specific parameter is used in the fake news detection on given event. The algorithm flow is same in the two stages and we use event adaption stage as an example to illustrate this procedure. 

Our input includes handful instances as context set $\{\mathbf{X}_{e}^s, \mathbf{Y}_{e}^s\} \setminus \{x^s_{e, i}, y^s_{e,i}\}$ and $x_{e,i}^s$ as  target data. We first feed $\mathbf{X}_{e}^s \setminus \{x^s_{e, i}\}$ into feature extractor and get context feature representations $\mathbf{C}_e^s$. The context feature representations $\mathbf{C}_e^s$ is then concatenated with label embedding $\mathbf{vec}_e^s$ of $\mathbf{Y}_e^s$. In the target side, the target data $x_{e,i}^s$ is also fed into feature extractor to get representation as $\mathbf{h}_{e,i}^s$. The aggragator component aggregates  $\mathbf{h}_{e,i}^s$, $\mathbf{C}_e^s$ and $\mathbf{vec}_e^s$ as introduced in section~\ref{section:hard_attention} to output context embedding $\mathbf{r}_{e,i}^s \in  \mathbb{R}^d$. Then we  concatenate $\mathbf{r}_{e,i}^s$ with target feature vector $\mathbf{h}_{e,i}^s \in  \mathbb{R}^d$.  The concatenated feature goes through the detector which is consisted of a fully connected layer to output a vector $\mathbf{o}_{e,i}^s$. The similarity scores between $\mathbf{o}_{e,i}^s$ and $\mathbf{vec(fake)}$, $\mathbf{vec(real)}$ are calculated  according to Eq.~\ref{eq:similiraty_fake} and Eq.~\ref{eq:similiraty_real} respectively. In the end, the similarity scores are mapped to probability values for fake news detection via softamax operation.

\section{Experiments}

In this section, we introduce the datasets used in the experiments, present the compared fake news detection models, validate the effectiveness and explore some insights of the proposed framework.

\subsection{Datasets}
To fairly evaluate the performance of the proposed model, we conduct experiments on datasets collected from two real-world social media datasets, namely Twitter and Weibo. The detailed description of the datasets are given below:

\begin{table}[htb]
  \caption{The Statistics of the Datasets.}
  \vspace{-0.1in}
\centering
  \begin{tabular}{|c|c|c|}
    \hline
    &Twitter & Weibo \\
     \hline
     \# of fake News &6,934 &4,050    \\ 
     \hline
     \# of real News &5,683 &3,558 \\  
  \hline
  \# of images &514& 7,606   \\ 
     \hline
\end{tabular}
    \label{tab:stat}
    \vspace{-0.1in}
\end{table}

\begin{table*}[!htb]
\caption{The performance comparison of models for fake news detection on the Twitter and Weibo datasets under 5-shot and 10-shot settings. Accuracy and F1 score of models are followed by standard deviation.  The percentage improvement ($\uparrow$) of MetaFEND over the best baseline per setting is in the last row. EANN, CNP, ANP, MAML, Meta-SGD and MetaFEND share the same feature extractor as the backbone.}
 \vspace{-0.1in}
\centering
\resizebox{0.9\textwidth}{!}{
\begin{tabular}{c|c|ccc|ccc|ccc|cc}
\toprule
&\multicolumn{5}{c}{Twitter} && \multicolumn{5}{c}{Weibo}  \\
\cline{2-6} \cline{8-12}
\multirow{2}{*}{Method}&\multicolumn{2}{c}{5-Shot} && \multicolumn{2}{c}{10-Shot}&&\multicolumn{2}{c}{5-Shot} && \multicolumn{2}{c}{10-Shot}\\
\cline{2-6} \cline{8-12}
&   Accuracy &   F1 Score   &&Accuracy &   F1 Score && Accuracy &   F1 Score &&Accuracy &   F1 Score \\
\midrule
VQA&73.62 $\pm$ 1.83 &76.69 $\pm$ 1.23 &&  73.49 $\pm$ 2.61  &74.69 $\pm$ 2.97 &&76.93 $\pm$ 0.71  &75.88 $\pm$ 0.45  && 77.80 $\pm$ 1.43  &76.36 $\pm$ 1.77\\
attRNN& 63.04 $\pm$ 2.09 &60.25 $\pm$ 4.63  && 63.14 $\pm$ 2.00 &56.60 $\pm$ 5.25 && 76.07 $\pm$ 1.63  &74.36 $\pm$ 2.96 &&78.09 $\pm$ 0.58  &77.69 $\pm$ 0.35  \\
EANN& 70.01 $\pm$ 3.58 &72.95 $\pm$ 2.86 && 70.56 $\pm$ 1.00 &67.77 $\pm$ 0.80 && 76.43 $\pm$ 0.84  &74.51 $\pm$ 0.56 && 77.49 $\pm$ 1.95  &76.56 $\pm$ 1.28 \\
CNP & 71.42 $\pm$ 2.58  &72.58 $\pm$ 3.57 && 72.47 $\pm$ 3.61  &72.11 $\pm$ 5.74 && 77.47 $\pm$ 5.19  &77.01 $\pm$ 4.66  && 78.81 $\pm$ 1.57 &78.07 $\pm$ 1.98\\
ANP& 77.08 $\pm$ 2.92  &79.65 $\pm$ 3.81  && 74.25 $\pm$ 0.76  &75.16 $\pm$ 1.27 && 77.85 $\pm$ 1.67 &76.00 $\pm$ 3.61  && 76.52 $\pm$ 1.84 &73.73 $\pm$ 2.78\\
MAML& 82.24 $\pm$ 1.54  &82.97 $\pm$ 1.76  && 85.22 $\pm$ 0.64 &84.98 $\pm$ 1.70 && 74.68 $\pm$ 0.75  &74.16 $\pm$ 0.33 &&75.87 $\pm$ 0.33  &73.41 $\pm$ 0.86  \\
Meta-SGD& 74.13 $\pm$ 2.31  &75.35 $\pm$ 2.56 && 74.63 $\pm$ 2.46 & 74.57 $\pm$ 2.74 && 71.73 $\pm$ 1.81 &69.51 $\pm$ 2.28   && 73.34 $\pm$ 2.35  &71.42 $\pm$ 2.80\\
\midrule
{MetaFEND} & \textbf{86.45 $\pm$ 1.83} &\textbf{86.21 $\pm$ 1.32}  && \textbf{88.79 $\pm$ 1.27} &  \textbf{88.66 $\pm$ 1.09} && \textbf{81.28 $\pm$ 0.75}  &\textbf{80.19 $\pm$ 1.27}  && \textbf{82.92 $\pm$ 0.13} & \textbf{82.37 $\pm$ 0.28} \\
(Improvement)& \textbf{($\uparrow$5.12\%)} & \textbf{($\uparrow$3.91\%)}& &\textbf{($\uparrow$4.19\%)}& \textbf{($\uparrow$4.33\%)}  && \textbf{($\uparrow$4.41\%)}& \textbf{($\uparrow$4.13\%)} &&\textbf{($\uparrow$5.22\%)} &\textbf{($\uparrow$5.51\%)}\\
\bottomrule
\end{tabular}
}
\vspace{-0.1in}
\label{tab: performance}
\end{table*}




The \textbf{Twitter dataset} is from MediaEval Verifying Multimedia Use benchmark~\citep{boididou2015verifying}, which is used in ~\citep{EANN,jin2017multimodal} for detecting fake content on Twitter. The \textbf{Weibo dataset}\footnote{https://github.com/yaqingwang/EANN-KDD18} is used in \citep{jin2017multimodal, EANN, qi2019exploiting} for detecting multi-modal fake news. The news events are included in the Twitter dataset and we follow the previous works~\citep{jin2017multimodal, EANN, qi2019exploiting} to obtain events on Weibo via  a single-pass clustering method~\citep{jin2014news}. In the two datasets above,  we only keep the events which are associated with more than 20 posts and randomly split the posts on same event into support and query data. To validate performance of the models on newly emergent events, we ensure that the training and testing sets do not contain any common event.  We adopt Accuracy and F1 Score as evaluation metrics. These two datasets cover diverse news events and thus can be used as good test-grounds for evaluation of fake news detection on heterogeneous events.

\subsection{Baselines}
To validate the effectiveness of the proposed model, we choose baselines from multi-modal models and the few-shot learning models. For the multi-modal models, we fine-tune them on support set from events in the testing data for a fair comparison. 
In the experiments, we have the 5-shot and 10-shot settings. In our problem, 5-shot setting refers to that 5 labeled posts are provided as support set.

\noindent \textbf{Fine-tune models.}
All the multi-modal approaches take the information from multiple modalities into account, including {VQA}~\citep{antol2015vqa}, {att-RNN}~\citep{jin2017multimodal} and EANN~\citep{EANN}. In the fine-tune setting, the training data including labeled support data and labeled query data is used to train the baselines. In the testing stage, the trained models are first fine-tuned on the labeled support data of given event, and then make predictions for testing query data. (1) \textbf{VQA}~\citep{antol2015vqa}. Visual Question Answering (\textbf{VQA}) model aims to answer the questions based on the given images and is used as a baseline for multimodal fake news in ~\citep{jin2017multimodal}. (2) \textbf{att-RNN}~\citep{jin2017multimodal}. \textsf{att-RNN} is the state-of-the-art model for multi-modal fake news detection. It uses attention mechanism to fuse the textual, visual and social context features. In our experiments, we remove the part dealing with social context information, but the remaining parts are the same. (3) \textbf{EANN}~\citep{EANN}.
{EANN} is one of the state-of-the-art models for fake news detection.
It consists of three components: feature extractor, event discriminator and fake news detector. It captures shared features across different events of news to improve generlziation ability.

\noindent \textbf{Few-shot learning models.} We use
 {CNP}~\citep{cnp}, {ANP}~\citep{anp}, MAML~\citep{MAML}  and Meta-SGD~\citep{MetaSGD} as few-shot learning baselines. (1) \textbf{CNP}~\citep{cnp}.  Conditional neural process 
is the state-of-the-art model for few-shot learning. It combines neural network and gaussian process by using a small set of input-output pairs as context to output predication for given input of data. (2) \textbf{ANP}~\citep{anp}. Attentive neural process belongs to the family of neural process which outputs prediction based on concatenation of learned distribution of context, context features and given input. (3) \textbf{MAML}~\citep{MAML}. Model-aganostic Meta-learning  is a representative optimization-based meta-learning model. The mechanism of MAML is to learn a set of shared model parameters across different tasks which can rapidly learn novel task with a small set of labeled data. (4) \textbf{Meta-SGD}~\citep{MetaSGD}. Meta-SGD is one of the state-of-the-art meta learning method for few-shot learning setting. Besides a shared global initialized parameters as with MAML, it also learns step sizes and update direction during the training procedure.

The proposed model share the same feature extractor backbone with EANN, CNP, ANP, MAML, Meta-SGD to study the effects of other designs in addition to benefits of the feature extractor backbone.

\noindent\textbf{Implementations} In the proposed model, the 300 dimensional FastText pre-trained word-embedding weights~\citep{fasttext} are used to initialize the parameters of the embedding layer. The window size of filters varies from 1 to 5 for textual CNN extractor. The hidden size $d_{f}$ of the fully connected layer in textual and visual extractor and dimension $d$ are set as 16 which is searched from options \{8, 16, 32, 64\}.  $\tau$ decays from $1$ to $0.5$ as the suggested way in~\citep{gumbel}. The gradient update step is set to 1 an inner learning rate $\beta$ is set to 0.1 for fine-tune models: MAML, Meta-SGD and our proposed framework MetaFEND. We implement all the deep learning baselines and the proposed framework with PyTorch 1.2 using NVIDIA Titan Xp GPU. For training models, we use Adam~\citep{adam} in the default setting. The learning rate $\alpha$ is 0.001.  We use mini-batch size of 10 and training epochs of 400.

\subsection{Performance Comparison}




Table~\ref{tab: performance} shows the performance of different approaches on the Twitter and Weibo datasets. We can observe that the proposed framework MetaFEND achieves the best results in terms of most of the evaluation metrics in both 5-shot and 10-shot settings. 

\noindent \textbf{Twitter.}
On the Twitter dataset in 5-shot setting, compared with CNP, ANP incorporates the attention mechanism and hence can achieve more informative context information.  Due to the heterogeneity of events, it is not easy for Meta-SGD to learn a shareable learning directions and step size across all events. Thus, Meta-SGD's performance is lower than MAML's in terms of accuracy. Compared with all the baselines, MetaFEND achieves the best performance in terms of most the metrics. Our proposed model inherits the advantages of MAML to learn a set of parameters which can rapidly learn to detect fake news with a small support set. Moreover,  MetaFEND can use the support data as conditioning set explicitly to better capture the uncertainty of events and thus it is able to achieve more than 5\% improvement compared with MAML in terms of accuracy.  
In the 10-shot setting, as the size of give support data increases, the soft attention mechanism of ANP unavoidably incorporates the irrelevant data points. In contrast, the proposed model MetaFEND employs the hard-attention mechanism to trim irrelevant data points from context set and  significantly outperforms all the baselines in terms of all the metrics.

\noindent \textbf{Weibo.}
Compared with the Twitter data, the Weibo dataset has different characteristics.  On the Weibo dataset, most of the posts are associated with different images. Thus, we can evaluate the performance of models under the circumstance where support datasets do not include direct clues with query set.  As EANN tends to ignore event-specific features, it achieves the lowest accuracy among fine-tune models in 10-shot setting. For the few-shot models, ANP and CNP achieves better performance compared with gradient-based meta-learning methods MAML and Meta-SGD. This is because the parameter adaptation may not be effective when support data set and query set do not share the same patterns.  Compared with ANP in 5-shot setting, our proposed method MetaFEND achieves 4.39\% improvement in terms of accuracy and 5.51\% improvement in terms of F1 score. The reason is that our MetaFEND can learn a base parameter which can rapidly learn to use a few examples as reference information for fake news detection. Thus, our proposed model enjoys the benefits of neural process and  meta-learning model families.

\vspace{-0.1in}
\subsection{Ablation Study}
\label{subsection:ablation_study}

We show ablation study to analyze the role of Hard-Attention and label embedding components. 

\begin{figure}[hbt]
\vspace{-0.1in}
\subfloat[Attention]{
\begin{minipage}{.23\textwidth}
\includegraphics[height=1.2in]{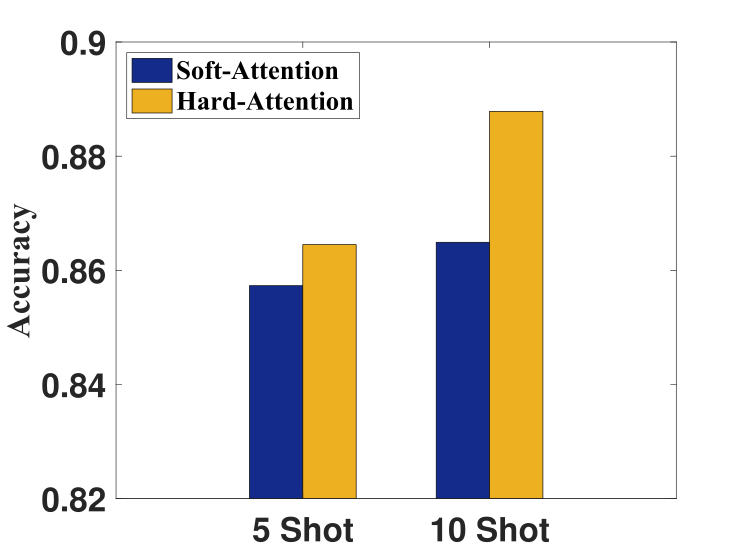}
\label{subfig:attention_acc}
\end{minipage}}
\subfloat[Label Embedding]{
\begin{minipage}{.23\textwidth}
\includegraphics[height=1.2in]{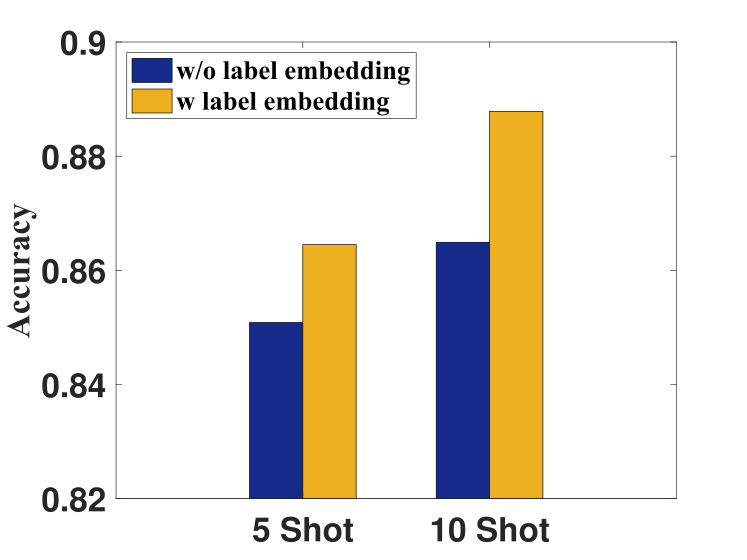}
\label{subfig:label_embedding_acc}
\end{minipage}}
\vspace{-0.1in}
 \caption{The ablation study about (a) Soft-Attention and Hard-Attention and (b) Label Embedding.}\label{fig:ablation_study}
\vspace{-0.2in}
\end{figure}

\noindent\textbf{Soft-Attention v.s. Hard-Attention.}
To intuitively illustrate the role of hard-attention mechanism in the proposed model,  we show ablation study by replacing hard-attention with soft-attention. Then we repeatedly run the new designed model on the Twitter dataset five times in 5-shot and 10-shot settings respectively and report the average of accuracy values. The results are show in the Figure~\ref{fig:ablation_study}. From Figure~\ref{subfig:attention_acc}, we can observe that accuracy scores of ``Hard-Attention'' in 5-shot and 10-shot settings are greater than those of ``Soft-Attention'' respectively.  As the number of support set increases, hard-attention mechanism does not have the limitation of soft-attention mechanism which unavoidably incorporates unrelated data points and significantly outperforms the soft-attention in terms of accuracy score. Thus, we can conclude that hard-attention mechanism can take effectively advantage of support set, and the superiority is more significant as we enlarge size of support set.

 \noindent\textbf{w/o Label Embedding  v.s. w/ Label Embedding.}
To analyze the role of label embedding in the proposed model, we design MetaFEND's corresponding reduced model by replacing label embedding with label value 0 or 1. Accordingly, we change the multiplication between output with label embedding to a binary-class fully connected layer to directly output the probabilities. Figure~\ref{subfig:label_embedding_acc}
 shows the results in terms of accuracy score. In Figure~\ref{subfig:label_embedding_acc}, ``w/o label embedding'' denotes that we remove the label embedding, and ``w label embedding'' denotes the original approach.  We can observe that the accuracy  score of ``w label embedding'' is greater than ``w/o label embedding'' in 5-shot and 10-shot settings,  demonstrating the effectiveness of label embedding


\subsection{Case Study}

\begin{figure}[hbt]
 \vspace{-0.1in}
\includegraphics[width=\linewidth]{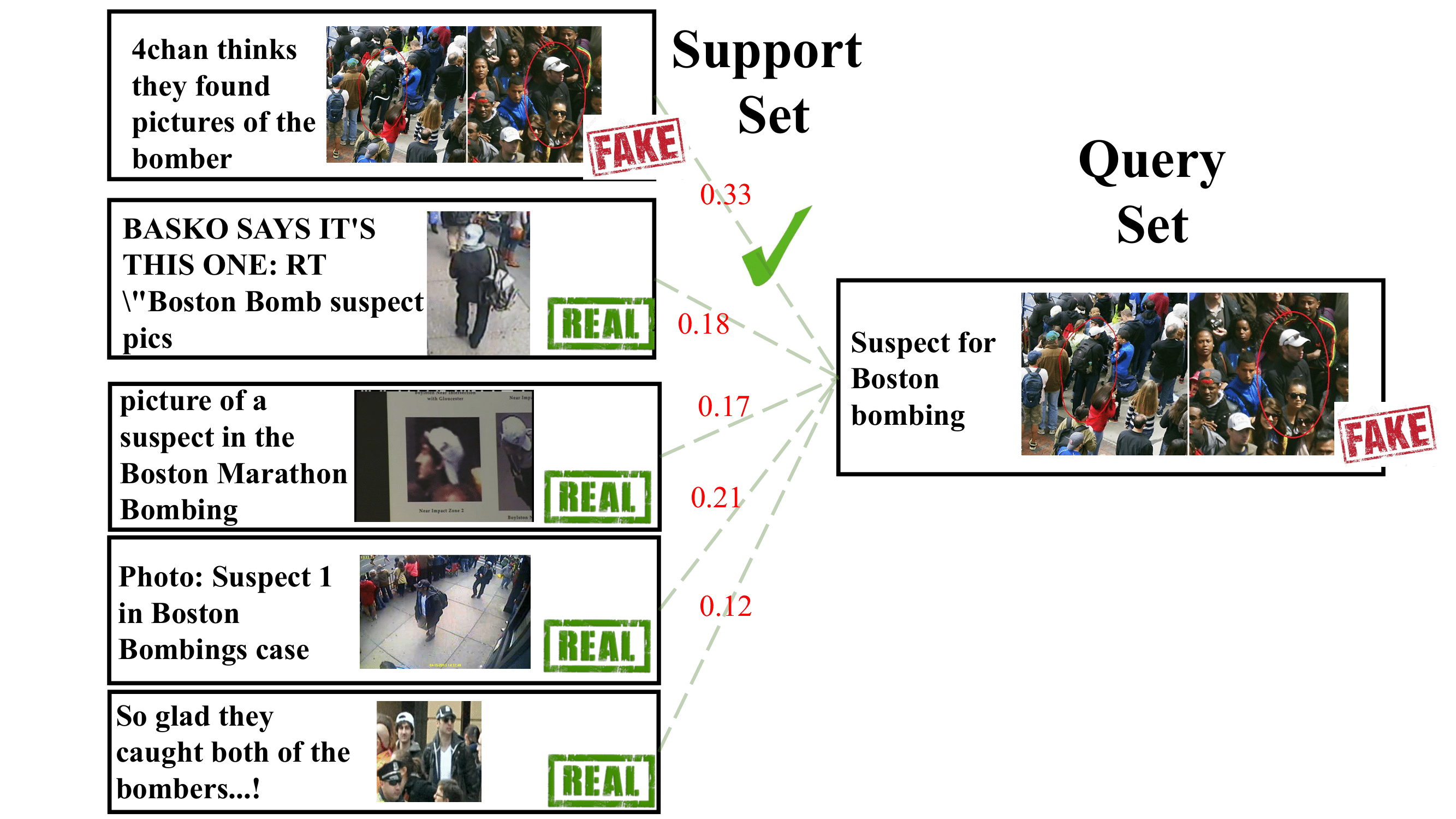}
 \caption{Fake news examples missed by Soft-Attention but spotted by Hard-Attention}\label{Fig:case_study}
\vspace{-0.05in}
\end{figure}

In order to illustrate the
challenges of emergent fake news detection and how our model handles challenges, we 
show one example in 5-shot learning setting as case study in Fig.~\ref{Fig:case_study}. As it can be observed, the four of five news examples in the support set are real news. Due to imbalanced class condition in the support set, it is difficult for Soft-Attention to provide correct prediction for news of interest in the query set. More specifically, Fig.~\ref{Fig:case_study} shows  the attention score values (red color) between examples in support set and query set based on multi-modal features. Although the first example with largest attention score value is most similar to news example in the query set, the majority of context information is from  the other four examples due to imbalanced class distribution. Such an imbalanced class distribution leads to incorrect prediction for Soft-Attention. The Hard-Attention mechanism can achieve correct result by focusing on the most similar sample in the support set. Through this example, we can also observe the necessity of event adaption stage. The posts and images for the same event are very similar and difficult to distinguish. Without event adaption stage, the model cannot capture informative clues to make correct predictions. 

\vspace{-0.1in}
\section{Related Work}
In this section, we briefly review the work related to the proposed model from fake news detection and few-shot learning.

\subsection{Fake News Detection} 

Many fake news detection algorithms try
to distinguish news according to their features, which can be extracted from  social context and news content. 
(1) \emph{Social context features} represent the user engagements of news on social media~\citep{shu2017fake} such as the number of followers, hash-tag (\#), propagation patterns~\citep{wu2015false} and retweets. 
However, social context features are very noisy, unstructured and labor intensive to collect. Especially, it cannot provide sufficient information for newly emerged events.  (2) \emph{Textual features} are statistical or semantic features extracted from text content of posts, which have been explored in many literatures of fake news detection~\citep{shu2017fake, gupta2014tweetcred,castillo2011information}. 
Unfortunately, linguistic patterns are not yet well understood, since they are highly dependent on specific events and corresponding domain knowledge~\citep{ruchansky2017csi}. 
To overcome this limitation, approaches like ~\citep{ma2016detecting, ma2018detect, ma2019detect, popat2018declare, lu2020gcan} propose to use deep learning models to identify fake news and have shown the significant improvements. (3) \emph{Visual features} have been shown to be an important indicator for fake news detection~\citep{jin2017novel,shu2017fake}. The basic features of attached images in the posts are explored in the work~\citep{gupta2012evaluating,ping2013review, jin2017novel}. 

In this paper, we consider multi-modal features when identifying fake news on social media. To tackle \emph{multi-modal fake news detection}, in~\citep{jin2017multimodal}, the authors propose a deep learning based fake news detection model, which extracts the multi-modal and social context features and fuses them by attention mechanism. To detect fake news on never-seen events, Wang et al.~\citep{EANN} propose an event-adversarial neural network (EANN) which can capture event-invariant features for fake news detection. However, EANN cannot take advantage of a small set of labeled data to further capture event specification and thus is not suited for our task.

 \vspace{-0.1in}
\subsection{Few-Shot Learning}

\textbf{Meta-learning} has long been proposed as a form of learning that would allow systems to systematically build up and re-use knowledge across different but related tasks~\citep{metalearning_survey,wang2020adaptive,wang2020automatic}.  MAML~\citep{MAML} is to learn model initialization parameters that are used to rapidly learn novel tasks with a small set of labeled data. Following this direction, besides initialization parameters, Meta-SGD~\citep{MetaSGD} learns step sizes and updates directions automatically in the training procedure. As tasks usually are different in the real setting, to handle task heterogeneity, HSML~\citep{hsml} customizes the global shared initialization to each cluster using a hierarchical clustering structure. The event heterogeneity is widely observed for fake news detection, where nonexistence of hierarchical relationship in news events makes this task more challenging.

\noindent\textbf{Neural process approaches}~\citep{np,anp, cnp}  combine stochastic process and neural network to handling task heterogeneity by conditioning on a context set. Conditional Neural Process (CNP)~\citep{cnp} and Neural Process (NP)~\citep{np} use neural networks to take input-output pairs of support set as conditioning for inference, incorporating task specific information.  However, these two works aggregate the context set by average or sum, ignoring different importance among context data samples and thereby leading to unsatisfactory performance. Attentive Neural Process (ANP)~\citep{anp} incorporates attention mechanism into Neural Process to alleviate such a issue. However, ANP still suffers from underfitting issue due to fixing parameters for different tasks. Additionally, ANP directly concatenates the label numeric values with feature representation, discarding the categorical characteristics of label information.

Different from existing works, our proposed framework maintains the parameter flexibility following the principle of meta-learning and inherits generalization ability to handle event heterogeneity from neural processes. Moreover, we incorporate label embedding component to handle categorical characteristics of label information and  utilize hard attention to extract most informative context information.  Thus, our proposed model enjoys the benefits of two model families without suffering their limitations. 

\vspace{-0.1in}
\section{Conclusions}
In this work, we study the problem of fake news detection on emergent events. The major challenge of fake news detection stems from newly emerged events on which existing approaches only showed unsatisfactory performance.  In order to address this issue, we propose a novel fake news detection framework, namely MetaFEND, which can rapidly learn to detect fake news for emergent events with a few labeled examples. The proposed framework can enjoy the benefits of meta-learning and neural process model families without suffering their own limitations.  Extensive experiments on two large scale datasets collected from popular social media platforms show that our proposed model MetaFEND outperforms the state-of-the-art models.
\vspace{-0.05in}
\section{Impact Statement}
Fake news can manipulate important public events and becomes a global concern. If the fake news detection algorithm can function as intended, it is beneficial to prevent the spread of fake news in the early stage and correspondingly many negative public events caused by fake news may be avoided. However, we are also aware that automatic detection may suppress the public discussion. The failure modes may lie in the negation cases: if someone tries to spot the fake news by citing false information contents, the automatic algorithm may not understand the logic behind the post and incorrectly identify it as fake news.  The bias may be unavoidable included in the dataset especially when the events are controversial or lacking a clear standard for annotation. Our proposed model explicitly uses the labeled sample as reference information and thus it is possible to replace the incorrect annotated support set by correct ones to correct the bias. To reduce harm brought by the automatic algorithm,  both technology and human review are needed and an effective user appeal system should be employed in case the incorrect detection happened.
\vspace{-0.1in}
\section*{Acknowledgment}
The authors thank the anonymous referees for their valuable comments and helpful suggestions. This work is supported in part by the US National Science Foundation under grants NSF IIS-1553411 and IIS-1956017. Any opinions, findings, and conclusions or recommendations expressed in this material are those of the author(s) and do not necessarily reflect the views of the National Science Foundation.
\vspace{-0.1in}
\bibliographystyle{ACM-Reference-Format}
\bibliography{sample-base}


\end{document}